\documentclass[10pt]{article}

\usepackage{cite}
 
\begin{document}

\null \hfill SPhT T01/25\\
\null \hfill math-ph/0103010\\
\vspace{0.5cm}
\begin{center}
{\Large \sf Higher--Order Corrections to Instantons}
\end{center}
\vspace{1cm}
\begin{center}
Ulrich~D.~Jentschura\dag\ and Jean Zinn-Justin\ddag
\end{center}
\vspace{0.2cm}
\begin{center}
\dag\ Laboratoire Kastler--Brossel, Unit\'{e} Mixte de Recherche
du CNRS No.~C8552,\\
Universit\'{e} Pierre et Marie Curie, Case 74, 4 pl.~Jussieu, 
F-75252 Paris Cedex 05, France\\[2ex]
\ddag\ Commissariat \`{a} l'\'{E}nergie Atomique,
Service de Physique Th\'{e}orique, \\
F-91191 Gif-Sur-Yvette Cedex, France\\[2ex]
{\bf Email:} ulj@spectro.jussieu.fr, zinn@spht.saclay.cea.fr
\end{center}
\vspace{1.3cm}
\begin{center}
\begin{minipage}{10.5cm}
{\underline{Abstract}}
The energy levels of the double-well potential receive, beyond
perturbation theory, contributions which are non-analytic in the coupling
strength; these are related to instanton effects. For example, the
separation between the energies of odd- and even-parity states is given
at leading order by the one-instanton contribution. However to determine
the energies more accurately multi-instanton configurations have also to
be taken into account. We investigate here the two-instanton
contributions. First we calculate analytically higher-order corrections to
multi-instanton effects. We then verify that the difference betweeen
numerically determined energy eigenvalues, and the generalized Borel sum
of the perturbation series can be described to very high accuracy by
two-instanton contributions. We also calculate higher-order corrections to
the leading factorial growth of the perturbative coefficients and show
that these are consistent with analytic results for the two-instanton
effect and with exact data for the first 200 perturbative coefficients.   
\end{minipage}
\end{center}
\vspace{1.3cm}

Asymptotic problems and properties
\newpage

In theories with degenerate minima, the energy eigenvalues
of the states at nonvanishing coupling $g \neq 0$ cannot 
{\em in principle} be obtained by analytic continuation from the 
unperturbed situation at vanishing coupling $g=0$ because a potential with 
degenerate minima introduces a degeneracy in the spectrum: for any
one unperturbed state, two states emerge when the perturbation
is switched on. These two states are seperated 
by an energy shift which is nonperturbative and nonanalytic in the coupling,
i.e.~vanishing to any order in perturbation theory.
Therefore, the two states are described by the same 
perturbation series and yet differ in their energy by instanton
contributions. Specifically, we consider
the case of the double-well potential with the hamiltonian
\begin{equation}
H = - \frac{g}{2} \, \frac{\partial^2}{\partial q^2} + \frac{1}{g} \, V(q)\,
\quad V(q) = \frac{1}{2}\,q^2\,(1 - q)^2\,.
\end{equation}
It has been conjectured~\cite{ZJ1981jmp,ZJ1981npb,ZJ1983npb,ZJ1984jmp} 
that an asymptotic expansion
for the energy eigenvalue can be obtained by finding a solution
to the equation
\begin{equation}
\label{quantization}
\frac{1}{\sqrt{2\pi}} \, \Gamma\left( \frac{1}{2} - D(E,g) \right) \,
\left(- \frac{2}{g} \right)^{D(E,g)} \, \exp[ -A(E,g)/2 ] = \pm {\rm i}\,,
\end{equation}
which can be understood as a modified Bohr-Sommerfeld quantization
condition. 
The plus and minus signs apply to even- and odd-parity
states, respectively. 
The conjecture (\ref{quantization}), whose validity has been
proven in~\cite{DeDi1991}, 
has found a natural explanation in
the framework of Ecalle's theory of resurgent 
functions~\cite{Ph1988,CaNoPh1993,Bo1994}.          
The functions $D(E,g)$, $A(E,g)$ constitute power series
in both variables. The function $D(E,g)$ describes the perturbative 
expansion; its evaluation is discussed in~\cite{ZJ1981jmp,ZJ1984jmp}. 
The first terms read
\begin{equation}
D(E,g) = E + g\,\left(3 \, E^2 + \frac{1}{4} \right) +
  g^2 \, \left(35 \, E^3 + \frac{25}{4} \, E \right) +
  \mathcal{O}(g^2) \,.
\end{equation}
The ground and the first excited state are both described
by the same perturbation series which can be found by
inverting the equation $D(E,g) = 1/2$. When the energy is expressed
in terms of the naive perturbation series in $g$,
the function $D(E,g)$ then vanishes in any order of perturbation
theory, i.e.~in all orders in $g$. For a general state, 
$D(E,g) = N + 1/2$ where $N$ is the quantum number of the unperturbed 
state which is a harmonic oscillator eigenstate.
The function $A(E,g)$ essentially describes instanton 
contributions~\cite{ZJ1984jmp}; its first terms read
\begin{equation}
A(E,g) = \frac{1}{3\,g} + g\,\left(17 \, E^2 + \frac{19}{12} \right) +
  g^2 \, \left(227 \, E^3 + \frac{187}{4} \, E \right) +
  \mathcal{O}(g^2) \,.
\end{equation}
A solution to the equation (\ref{quantization}) can be found by 
systematically expanding the energy eigenvalue $E(g)$ in powers of
$g$ and in the two quantities
\[
\lambda(g) = \ln\left( - \frac{2}{g} \right) \quad
\mbox{and} \quad \xi(g) = \frac{\exp[-1/(6 g)]}{\sqrt{\pi g}}\,.
\]
Terms of order $\xi(g)^n$ belong to the $n$-instanton contribution.
The energy eigenvalue for nonvanishing perturbation $g\neq 0$ can be 
described by two quantum numbers: the unperturbed quantum number $N$
and the positive or negative parity of the state. We have (the upper
index denotes the instanton order)
\begin{equation}
\label{ExpansionIntoInstantons}
E_{N,\pm}(g) = \sum_{n=0}^{\infty} E^{(n)}_{N,\pm}(g)
\end{equation}
where the perturbation series (zero-instanton contribution) is given as
\begin{equation}
\label{PerturbationSeries}
E^{(0)}_{N,\pm}(g) = \sum_{K=0}^{\infty} E^{(0)}_{N,K} \, g^K \,,
\end{equation}
where the right-hand side is parity independent. For 
$n > 0$, the instanton contribution reads
\begin{equation}
\label{InstantonContribution}
E^{(n)}_{N,\pm}(g) = \left( \frac{2}{g} \right)^{N n} \,
\xi(g)^n \, \sum_{k = 0}^{n-1} \lambda(g)^k \sum_{l=0}
\epsilon^{(N,\pm)}_{nkl} \, g^l \,.
\end{equation}
The lower indices $n$, $k$ and $l$ of
the $\epsilon$ coefficients denote the instanton
order, the power of the logarithm and the power of $g$, respectively.
The results relevant for the current investigation read,
\begin{eqnarray}
\label{Results}
\epsilon^{(0,+)}_{100} &=& - \epsilon^{(0,-)}_{100} = -1 \,, \quad
\epsilon^{(0,+)}_{101} = - \epsilon^{(0,-)}_{101} = \frac{71}{12} \,,
\nonumber\\[2ex]
\epsilon^{(0,+)}_{101} &=& - \epsilon^{(0,-)}_{101} = \frac{6299}{288} 
\,, \quad
\epsilon^{(0,+)}_{210} = \epsilon^{(0,-)}_{210} = 1 \,,
\nonumber\\[2ex]
\epsilon^{(0,+)}_{211} &=& \epsilon^{(0,-)}_{211} = -\frac{53}{6} 
\,, \quad
\epsilon^{(0,+)}_{212} = \epsilon^{(0,-)}_{212} = -\frac{1277}{72} \,,
\nonumber\\[2ex]
\epsilon^{(0,+)}_{200} &=& \epsilon^{(0,-)}_{200} = \gamma 
\,, \quad
\epsilon^{(0,+)}_{201} = 
\epsilon^{(0,-)}_{201} = -\frac{23}{2} - \frac{53}{6} \, \gamma \,,
\nonumber\\[2ex]
\epsilon^{(0,+)}_{202} &=& 
\epsilon^{(0,-)}_{202} = \frac{13}{12} - \frac{1277}{72} \, \gamma \,,
\end{eqnarray}
where $\gamma = 0.57221\dots$ is Euler's constant.
Odd-instanton contributions have opposite sign for opposite-parity
states and are responsible, in particular, for the energy difference 
of the ground state with quantum numbers
$(0,+)$ and the first excited state 
with quantum numbers $(0,-)$.
The dominant contribution to the seperation of the two
lowest energy levels is given by the one-instanton contribution:
\begin{equation}
\label{Separation}
E_{0,-}(g) - E_{0,+}(g) \sim 2 \, \xi(g) \, 
\left(1 - \frac{71}{12} \, g - \frac{6299}{288} \, g^2 +
\mathcal{O}(g^3)\right) + \mathcal{O}(\xi(g)^3)\,.
\end{equation}
By contrast, even-instanton contributions have like sign for 
opposite-parity 
states and are responsible, in particular, for the displacement
of the mean value $(1/2) \, [E_{0,-}(g) + E_{0,+}(g)]$ from the 
value of the generalized Borel sum of the perturbation series
$\mathcal{B}\left(\sum_{K=0}^{\infty} E^{(0)}_{0,K} \, g^K\right)$
(for the evaluation of the generalized Borel sum of a nonalternating
divergent series see for example Sec.~VI and Table III
of~\cite{Je2000prahep}).
The dominant contribution to the displacement comes from 
the two-instanton effect, and we have
\begin{eqnarray}
\label{Displacement}
\lefteqn{\frac{1}{2} \, [E_{0,-}(g) + E_{0,+}(g)] -
\mathrm{Re} \left\{
\mathcal{B}\left(\sum_{K=0}^{\infty} E^{(0)}_{0,K} \, g^K\right)
\right\} \sim 
\xi(g)^2 \, \left\{ \ln\left(\frac{2 \mathrm{e}^\gamma}{g}\right) 
\right.}\nonumber\\[2ex]
&& \left. + g \left[ - \frac{53}{6}  
\ln\left(\frac{2 \mathrm{e}^\gamma}{g}\right) \! - \! \frac{23}{2} \right]
+ g^2 \left[ - \frac{1277}{72} 
\ln\left(\frac{2 \mathrm{e}^\gamma}{g}\right) \! + \!
\frac{13}{12} \right]  + \mathcal{O}(g^3 \ln(g)) \right\} 
+ \mathcal{O}(\xi(g)^4)\,. 
\end{eqnarray}
The function~\cite{ZJ1981npb}
\begin{equation}
\label{DefinitionOfDelta}
\Delta(g) = 4 \, \frac{
\frac{\displaystyle 1}{\displaystyle 2} \, [E_{0,-}(g) + E_{0,+}(g)] - 
\mathrm{Re} \left\{
\mathcal{B}\left(\sum_{K=0}^{\infty} E^{(0)}_{0,K} \, g^K\right)\right\}}
{\left[E_{0,-}(g) - E_{0,+}(g)\right]^2 \, \ln\left(
\frac{\displaystyle 2 \mathrm{e}^\gamma}{\displaystyle g}\right)}\,.
\end{equation}
relates the multi-instanton contributions to the energy eigenvalues,
which can be evaluated numerically,
and to the (generalized) Borel sum of the perturbation series
which is evaluated by analytic continuation of the 
integration path into the complex plane (see~\cite{ZJ1996}). 
The calculation of $\Delta(g)$ at small coupling is 
problematic because of severe numerical cancellations.
From the equations (\ref{Separation}), (\ref{Displacement}) and
(\ref{DefinitionOfDelta}), we obtain the following asymptotics for 
$\Delta(g)$,
\begin{eqnarray}
\label{DeltaAsymptotics}
\Delta(g) \sim 1 + g\,\left[ \frac{71}{6} +
\left( -\frac{53}{6} \, 
\ln\left(\frac{2\mathrm{e}^\gamma}{g}\right) - \frac{23}{2} \right)
\bigg/
\ln\left(\frac{2\mathrm{e}^\gamma}{g}\right) \right] \nonumber\\[2ex]
+ g^2 \,\left[\frac{10711}{72} +
\left( \frac{1277}{72} \,
\ln\left(\frac{2\mathrm{e}^\gamma}{g}\right) - \frac{13}{12} \right)
\bigg/ \ln\left(\frac{2\mathrm{e}^\gamma}{g}\right)    
\right] + \mathcal{O}(g^3) \,.
\end{eqnarray}
If we additionally perform an expansion in inverse powers
of $\ln(2/g)$ and keep only the first few terms in 
$\{1/\ln(2/g)\}$ in each term in the $g$-expansion, the result reads
\begin{eqnarray}
\lefteqn{\Delta(g) \sim 1 + 3 g - \frac{23}{2} \frac{g}{\ln(2/g)} \,
\left[1 - \frac{\gamma}{\ln(2/g)} + \frac{\gamma^2}{\ln^2(2/g)} +
\mathcal{O}\left(\frac{1}{\ln^3(2/g)}\right) \right]}\nonumber\\[2ex] 
&& + \frac{53}{2} g^2 - 135 \, \frac{g^2}{\ln(2/g)} \,
\left[1 - \frac{\gamma}{\ln(2/g)} + \frac{\gamma^2}{\ln^2(2/g)}  + 
\mathcal{O}\left(\frac{1}{\ln^3(2/g)}\right) \right]
+ \mathcal{O}\left(g^3\right) \,.
\end{eqnarray}
The higher-order corrections, which are only
logarithmically suppressed with respect to
the leading terms $1 + 3g$, change the numerical values 
quite significantly,
even at small coupling. In Table~\ref{table1} we present numerical
results for the function $\Delta(g)$ at small coupling; these are
in agreement with the first few asymptotic terms listed in equation
(\ref{DeltaAsymptotics}) up to numerical accuracy. Of course, 
for strong coupling, significant deviations
from the leading asymptotics must be expected due to higher-order
effects; these are indeed
observed. For example, at $g=0.1$ the numerically
determined value reads $\Delta(0.1) = 0.87684(1)$ whereas the first
asymptotic terms given in equation (\ref{DeltaAsymptotics}) 
sum up to a numerical value of $0.86029$.  

%
% table1
%
\begin{table}[tbh]
\begin{center}
\begin{minipage}{15cm}
\begin{center}
\caption{\label{table1} Comparison of numerical values for
the function $\Delta(g)$ defined in equation (\ref{DefinitionOfDelta})
in the region of small coupling to values obtained by
calculating the first few terms in its asymptotic expansion 
given in (\ref{DeltaAsymptotics}).}
\vspace*{0.3cm}
\begin{tabular}{cr@{.}lr@{.}lr@{.}lr@{.}lr@{.}lr@{.}l%
r@{.}lr@{.}lr@{.}lr@{.}lr@{.}l}
\hline
\hline
\rule[-3mm]{0mm}{8mm} coupling $g$ &
 $0$ & $005$ &
 $0$ & $006$ &
 $0$ & $007$ &
 $0$ & $008$ &
 $0$ & $009$ &
 $0$ & $010$ \\
\hline
$\Delta(g)$ num. &
\rule[-3mm]{0mm}{8mm}
 $1$ & $0063(5)$ &
 $1$ & $0075(5)$ &
 $1$ & $00832(5)$ &
 $1$ & $00919(5)$ &
 $1$ & $00998(5)$ &
 $1$ & $01078(5)$ \\
$\Delta(g)$ asymp. &
\rule[-3mm]{0mm}{8mm}
 $1$ & $00640$ &
 $1$ & $00739$ &
 $1$ & $00832$ &
 $1$ & $00919$ &
 $1$ & $01001$ &
 $1$ & $01078$ \\             
\hline
\hline
\end{tabular}
\end{center}
\end{minipage}
\end{center}
\end{table}

The higher-order corrections to the two-instanton effect are related 
to the corrections to the leading factorial growth of the perturbative
coefficients. This can be seen by expressing that the imaginary part 
of the perturbation series, when continued analytically from 
negative to positive coupling, has to cancel with the imaginary
part of the two-instanton contribution which is generated by the 
logarithms $\ln(-2/g)$. The corrections of order $g\,\ln(-2/g)$
and $g^2\,\ln(-2/g)$ yield the $1/K$-- and $1/K^2$--corrections
to the leading factorial growth of the perturbative coefficients.
From the results for $\epsilon^{(0,\pm)}_{21j}$ ($j = 0,1,2$) given in 
equation (\ref{Results}), we obtain
\begin{equation}
\label{Corrections}
E^{0}_{0,K} \sim - \frac{3^{K+1} \, K!}{\pi} \left[ 1 -
\frac{53}{18} \, \frac{1}{K} -
\frac{1277}{648} \, \frac{1}{K^2} + \mathcal{O}\left(\frac{1}{K^3}\right)
\right]\,.
\end{equation}
The analytic results should be checked against explicit values
of the perturbative coefficients.
We have determined the first 200 perturbative coefficents 
$E^{(0)}_{0,K}$ ($K = 0,\dots, 200$) of the perturbation in 
the form of rational numbers, i.e.~to formally infinite
numerical accuracy. This allows to verify the 
$1/K$-- and $1/K^2$--corrections to the leading factorial growth in
equation (\ref{Corrections}) to high accuracy, for example by 
employing Richardson extrapolation~\cite{Ri1927}.
Using the 160th through the 200th perturbation coefficient
as input data for the Richardson algorithm,
the coefficients of the leading, of the $1/K$-subleading and of the 
$1/K^2$ suppressed corrections are found to be consistent with the 
analytic results given in equation (\ref{Corrections}) up to a relative
numerical accuracy of $10^{-26}$, $10^{-23}$ and $10^{-20}$, 
respectively. 
For completeness, we give here the numerical
values of the 198th through the 200th perturbative coefficients,
to 30 decimals. These read:
\begin{eqnarray}
E^{(0)}_{0,198} = 5.50117\,76962\,88587\,93527\,75694\,38632 \times
  10^{464}\,, \nonumber\\
E^{(0)}_{0,199} = 3.28445\,39841\,65780\,00616\,21912\,32835 \times
  10^{467}\,, \nonumber\\
E^{(0)}_{0,200} = 1.97082\,14193\,09543\,76979\,53006\,07410 \times
  10^{470}\,.
\end{eqnarray}
Values for all 200 coefficients will be presented elsewhere.

It is an interesting consequence of the expansion 
(\ref{ExpansionIntoInstantons}) that the energy difference 
$(E_{0,-} - E_{0,+})$, at small coupling, is described 
to high accuracy by the one-instanton contribution ($n=1$ in 
equation (\ref{InstantonContribution})).
For $g=0.001$, we obtain to 180 decimals, 
\begin{eqnarray}
\label{Result0plus}
E_{0,+}(0.001) &=& 
0.49899\,54548\,62109\,17168\,91308\,39481\,92163\,68209\,47240\,
\nonumber\\
& & \;\;\; 20809\,66532\,93278\,69722\,01391\,
                            \underline{15135\,28505\,38294\,45798}\,
\nonumber\\
& & \;\;\; 
\underline{45759\,95999\,06739\,55175\,84722\,67802\,81306\,96906\,01325}\,
\nonumber\\
& & \;\;\;
\underline{25943\,77289\,94365\,88255\,24440\,17437\,12789\,27978\,99793}\,,
\end{eqnarray}
whereas
\begin{eqnarray}   
\label{Result0minus}
E_{0,-}(0.001) &=&
0.49899\,54548\,62109\,17168\,91308\,39481\,92163\,68209\,47240\,
\nonumber\\
& & \;\;\; 20809\,66532\,93278\,69722\,01391\,
                     \underline{29839\,92959\,55803\,70812}\,
\nonumber\\
& & \;\;\; 
\underline{27749\,92448\,48259\,36743\,64757\,68328\,84835\,35511\,34663}\,
\nonumber\\ 
& & \;\;\; 
\underline{06309\,82331\,51885\,23308\,08622\,84780\,52722\,10103\,67282}\,.
\end{eqnarray}
Decimals which differ in the two energy levels are underlined.
The results have been obtained by lattice extrapolation using a
modified Richardson algorithm which is constructed according to 
ideas outlined in~\cite{We1989}. Calculations were performed on 
IBM RISC/6000 workstations while making extensive use of 
multiprecision libraries~\cite{Ba1990tech,Ba1993,Ba1994tech}.
We define $\mathcal{P}_M(g)$
as the $M$th partial sum of the one-instanton contribution
$E^{(1)}_{0,-}(g) - E^{(1)}_{0,+}(g)$,
\begin{equation}
\mathcal{P}_M(g) =
2 \, \xi(g) \, \sum_{j = 0}^{M} \epsilon^{(0,-)}_{10j} \, g^j \,.
\end{equation}               
Using exact rational expressions for the 
coefficients $\epsilon^{(0,-)}_{10j}$ ($j \leq 141$),
we obtain
\begin{eqnarray}
& & \mathcal{P}_{140}(0.001) \times 10^{71} = \nonumber\\[1ex]
& & \;    1.47046\,44541\,75092\,50138\,19899\,%
64494\,15198\,15678\,00350\,05260\,35283\,,\nonumber\\
& & \;\;\;\; 86053\,33378\,03660\,50415\,75193\,%
50528\,41826\,73433\,99328\,21246\,74888\,\; \\[3ex]
& & \mathcal{P}_{141}(0.001) \times 10^{71} = \nonumber\\[1ex]
& & \; 1.47046\,44541\,75092\,50138\,19899\,%
64494\,15198\,15678\,00350\,05260\,35283\,,\nonumber\\
& & \;\;\;\; 86053\,33378\,03660\,50415\,75193\,%
50528\,41826\,73433\,99328\,21246\,74887\,.    
\end{eqnarray}
These values are in excellent agreement with the 
numerically determined energy difference (see the 
results presented above in equations (\ref{Result0plus}) and
(\ref{Result0minus}))
\begin{eqnarray}
& & [E_{0,-}(0.001) - E_{0,+}(0.001)] \times 10^{71} = \nonumber\\[1ex]
& & \; 1.47046\,44541\,75092\,50138\,19899\,%
64494\,15198\,15678\,00350\,05260\,35283\,,\nonumber\\
& & \;\;\;\; 86053\,33378\,03660\,50415\,75193\,%
50528\,41826\,73433\,99328\,21246\,74887\,.       
\end{eqnarray}
The first 70 decimals in equations (\ref{Result0plus}) and
(\ref{Result0minus}) are the same because the one-instanton 
contribution is of the order of $1.4\times 10^{-71}$.
The accuracy to which the one-instanton contribution describes
the energy difference $E_{0,-}(0.001) - E_{0,+}(0.001)$ is limited
by the three-instanton effect which for $g=0.001$ is of the order of
$8\times 10^{-212}$. Note that the two-instanton effect 
(which for $g=0.001$ is of the
order of $4\times 10^{-142}$) does not limit the accuracy to which 
the one-instanton contribution describes the energy {\em difference}
because it has the same sign and equal magnitude
for opposite-parity states with the same unperturbed quantum
number $N$.

We have demonstrated that the behavior of the 
characteristic function $\Delta(g)$
defined in equation (\ref{DefinitionOfDelta}) 
at small coupling is consistent with higher-order corrections to the 
one- and two-instanton contributions, specifically with the 
instanton expansion of the energy levels
governed by the equations
(\ref{ExpansionIntoInstantons}) and (\ref{InstantonContribution}), 
with the assumption that the instanton contributions
given by equation (\ref{InstantonContribution}) should
be Borel summed, with the 
explicit results for the higher-order coefficients
listed in (\ref{Results}) and the analytically derived 
asymptotics for the function $\Delta(g)$ 
given in equation (\ref{DeltaAsymptotics}).
The corrections of relative order $1/K^m$ 
to the leading factorial growth
of the perturbative coefficients -- see equation (\ref{Corrections}) --
are consistent with the analytically evaluated $g^m\,\ln(-2/g)$--corrections
to the two-instanton effect and with the explicit values for 
the first 200 terms in the perturbation series (\ref{PerturbationSeries}).
The nonperturbative energy difference $E_{0,-}(g) - E_{0,+}(g)$
at small coupling $g$ is described, to high accuracy, by the
one-instanton contribution only.

%
% Acknowledgements
%
\section*{Acknowledgements}

U.D.J. acknowledges support from the Deutscher
Akademischer Austauschdienst (DAAD).

\end{document}